\documentclass[10pt,a4paper,twocolumn,showpacs,amsmath,pra,amssymb,english,nofootinbib]{revtex4}
\usepackage[T1]{fontenc}
\usepackage[latin9]{inputenc}
\usepackage{float}
\usepackage{graphicx}
\usepackage{amssymb}
\usepackage{esint}
\usepackage{times}

\makeatletter

\usepackage{babel}
\makeatother

\begin{document}

\title{Transition region properties of a trapped quasi-two-dimensional degenerate Bose gas}

\author{R.~N.~Bisset and P.~B.~Blakie}
\affiliation{Department of Physics, Jack Dodd Centre for Quantum Technology, University of Otago, P.O. Box 56, Dunedin, 9016 New Zealand}


\date{\today}
\begin{abstract}
The classical field simulation technique is used to study a trapped quasi-two dimensional Bose gas. We calculate the central curvature of the system density and fluctuations of the condensate mode in the degenerate regime. These results provide new understanding of the system behavior in the region of the superfluid transition.\end{abstract}

\maketitle
\section{Introduction}

Evidence for the Berezinskii-Kosterlitz-Thouless (BKT) superfluid transition in a dilute quasi-two dimensional (quasi-2D) Bose gas was first reported by the ENS group \cite{Hadzibabic2006}. In that work the algebraic decay in first order correlations, inferred from an interference procedure, was used to identify the transition point. Subsequent work suggested that the transition could also be identified by bimodality in the system density after expansion \cite{Kruger2007}. In contrast, recent experiments at NIST  \cite{Clade2009}  found that the onset of bimodality and superfluidity are distinct, with bimodality occurring at a higher temperature. 

Meanfield methods are inapplicable to the regime of interest and theoretical understanding of this system is just beginning to emerge with the development of classical field (c-field) \cite{Simula2006} and quantum Monte Carlo (QMC) \cite{Holzmann2008}
methods for simulating the trapped quasi-2D Bose gas. 
Previous results obtained with the c-field technique have found the onset of bimodality and BKT superfluidity to be distinct, in qualitative agreement with the NIST observations \cite{Bisset2009}. We also note work on a semiclassical field method \cite{Giorgetti2007a}, which was applied to studying thermal vortices in a homogeneous 2D Bose gas.

Many important questions remain to be answered for a complete understanding of this system, particularly in the transition region. Here we address two important issues: First, we examine the use of the central density curvature ($\kappa$) as a means to identify the transition point. This quantity was originally proposed by Holzmann and Krauth \cite{Holzmann2008} who showed, using QMC calculations, that $\kappa$ was peaked in the vicinity of the superfluid transition. 
However, their study contained only a small number of results across the transition region, and was unable to resolve any distinction between bimodality and the onset of BKT superfluidity. 
The efficiency of the c-field method allows us to characterize the behavior of $\kappa$ with fine resolution across the transition region. In addition to providing us with a better understanding of this observable in the region of current interest, it also provides a useful qualitative comparison of the c-field and QMC  \cite{Holzmann2008} techniques.
Second, we present results characterizing the properties of the system in the temperature range where the gas is bimodal but has not yet attained BKT superfluidity. In previous work  \cite{Bisset2009} we showed that the system has a small but finite condensate fraction in this temperature range. However, in this regime the condensate appeared to fluctuate strongly and easily be penetrated by single vortices (also see \cite{Clade2009}). We apply a newly developed technique \cite{Bezett2009b} to sample the condensate number distribution and reveal its fluctuations. These new results shed important light on the physics of the trapped quasi-two dimensional Bose gas, and suggest some new directions for experimental investigation.

\section{Formalism}
An ultra-cold Bose  gas is described by the Hamiltonian
\begin{equation}
\hat{H}=\int d^{3}\mathbf{x}\,\hat{\Psi}^{\dagger}(\mathbf{x})\left\{ H_{\rm{sp}}+\frac{2\pi a\hbar^{2}}{m}\hat{\Psi}^{\dagger}(\mathbf{x})\hat{\Psi}(\mathbf{x})\right\} \hat{\Psi}(\mathbf{x}),\label{eqH}
\end{equation}
where $m$ is the atomic mass, $a$ is the $s$-wave scattering length and the single-particle Hamiltonian is given by
\begin{equation}
H_{\rm{sp}}=\frac{p^{2}}{2m}+\frac{1}{2}m(\omega_x^2x^2+\omega_y^2y^2+\omega_z^2z^2).
\end{equation}
In the regime  where the trap frequencies satisfy $\omega_z \gg \omega_x,\omega_y$, motion in the  $z$-direction may be frozen out, resulting in a quasi-2D regime.
  Eq.~(\ref{eqH}) assumes three-dimensional scattering between the atoms which
requires that the $z$-confinement length, $a_z = \sqrt{\hbar/m\omega_z}$, satisfies  $a_z \gg a$ \cite{Petrov2000}.   These requirements are well-satisfied by experiments \cite{Stock2005,Hadzibabic2006,Kruger2007,Clade2009}.

Our simulation method is based on a c-field representation of the highly occupied low energy modes (i.e.~those below an appropriately chosen energy cutoff, see \cite{Blakie2007CR}) and a Hartree-Fock treatment of the remaining sparsely occupied (high energy) modes of the system. An advantage of this method is that the  critical modes are contained within the \emph{c-field} description and are  treated non-perturbatively. For details of this c-field theory refer to \cite{Davis2001,Blakie2005,Blakie2008} and for the specific application to the quasi-2D trapped Bose gas see \cite{Simula2006,Simula2008,Sato2009,Bisset2009}.

\section{Results}\label{Sec:Results}
We simulate $^{87}\rm{Rb}$ atoms in a cylindrically symmetric trap (i.e.~$\omega_{x,y}=\omega$), presenting results for two sets of parameters  in which the dimensionless 2D interaction strength, $\tilde{g} = \sqrt{8\pi}a/a_z$ \cite{Petrov2000}, takes the values $0.075$ and $0.107$. For comparison, the ENS experiment (with $^{87}$Rb) was in the regime where $\tilde{g}\approx0.13$ \cite{Hadzibabic2006} and for the NIST experiment  (with $^{23}$Na) $\tilde{g}\approx0.02$ \cite{Clade2009}.
In this paper the densities given are areal (i.e.~integrated over the $z$-direction) and are dependent on the 2D position vector $\mathbf{r}=(x,y)$.

\subsection{Degenerate components}\label{Sec:DegenCpts}
In previous work \cite{Bisset2009} we identified the various components contributing to the phase diagram. Here we briefly review the definitions of these components and summarize those results.
\paragraph{Quasicondensate}  is the component of the system with suppressed density fluctuations, defined by \cite{Prokofev2001}
\begin{equation}
n_{qc}(\mathbf{r})=\sqrt{2\langle\hat{n}(\mathbf{r})\rangle^{2}-
\langle\hat{n}(\mathbf{r})^{2}\rangle},\label{Eq:QuasiC}
\end{equation}
where $\hat{n}(\mathbf{r})$ is the density operator for the system. Note, a normal system with Gaussian fluctuations has $n_{qc} = 0$.  

\paragraph{Condensate/Coherence} is identified by
a macroscopic  eigenvalue  in  the one-body density matrix
$G(\mathbf{r},\mathbf{r}')=\langle\hat{\Psi}^{\dagger}(\mathbf{r})\hat{\Psi}(\mathbf{r}')\rangle$, 
\cite{Blakie2005,Bisset2009}. The condensate is expected to vanish for the uniform 2D gas in the thermodynamic limit, however our results show that it plays a role in the finite trapped samples realized in experiments (also see \cite{Bagnato1991,Holzmann2008}).
 \paragraph{Superfluid model of Holzmann and Krauth}
In Ref.~\cite{Holzmann2008}  a model for the superfluid component was proposed, based on a local density application of the uniform results \cite{Prokofev2001,Prokofev2002} to the trapped system.
This model predicts a superfluid component wherever the local density exceeds  the critical value,
\begin{equation}
n_{cr}=\lambda^{-2}\log\left(\frac{C}{\tilde{g}}\right),\label{Eq:peakdencond}
\end{equation}
with $C=380\pm3$ and $\lambda=h/\sqrt{2\pi mk_{B}T}$ \cite{Prokofev2001}. The temperature at which the peak phase space density satisfies the critical condition (\ref{Eq:peakdencond}) is denoted $T_{BKT}$ \cite{Holzmann2008,Holzmann2008A,Bisset2009,Bisset2009A}\footnote{We use the lowest axial mode areal density to identify $T_{BKT}$, whereas \cite{Holzmann2008A} use the total areal density, also see \cite{Bisset2009A}.}.
 
\begin{figure}
\centering
\includegraphics[width=3.2in]{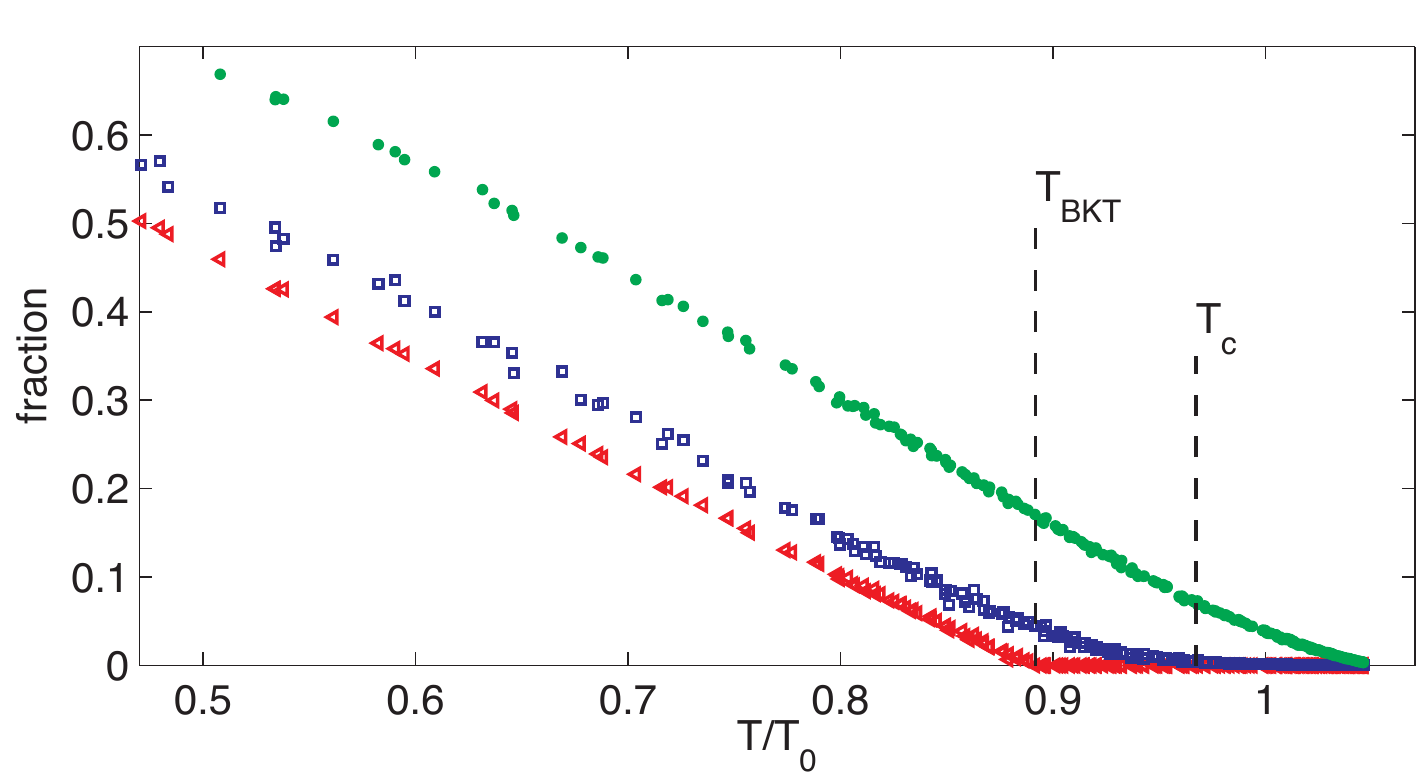}
\caption{\label{Fig:DegenFracs} Fraction of atoms in the quasicondensate (dots), condensate (squares) and superfluid (triangles) components as a function of reduced temperature. Figure from Ref.~\cite{Bisset2009}. Parameters: $(\omega_{x,y},\omega_z)=2\pi\times(9.4,1880)$ Hz, $N=13.5\times10^3$  atoms at $T_{BKT}=34.7$ nK, with $\tilde{g}=0.107$.}
\end{figure}

The results of Ref.~\cite{Bisset2009} for these components are summarized in Fig.~\ref{Fig:DegenFracs}, plotted against reduced temperature, $T^\prime=T/T_0$, where $T_0$ is the ideal gas condensation temperature\footnote{An improvement in our meanfield treatment of the incoherent region atoms has led to a small change in the fractions compared to those reported in \cite{Bisset2009}.}. 

These results clearly reveal that the quasicondensate, condensate and superfluid components are distinct in the quasi-2D system. Furthermore, we observe that, in order of decreasing temperature:
(i) The first manifestation of degeneracy is that density fluctuations are suppressed, i.e.~a quasicondensate forms (this begins to occur at $T'\gtrsim1$ in Fig.~\ref{Fig:DegenFracs}).
(ii) At $T_c$ a condensate appears and the momentum distribution of the system becomes bimodal. This crossover is a finite size effect, but occurs well-separated from the BKT transition for the typical experimental regime. We identify $T_c$ as where the largest eigenvalue of $G(\mathbf{r},\mathbf{r}')$  exceeds the second largest by a factor of 1.5, at which point momentum space bimodality is clearly apparent \cite{Bisset2009}.
(iii) At $T_{BKT}$ the peak phase space density is sufficiently high for BKT superfluidity to emerge in the system.
More details on these results and the simulation method are given in  Ref.~\cite{Bisset2009}.

\subsection{Central curvature of density distribution}\label{Sec:CentralCurve}
\begin{figure}
\centering
\includegraphics[width=3.1in]{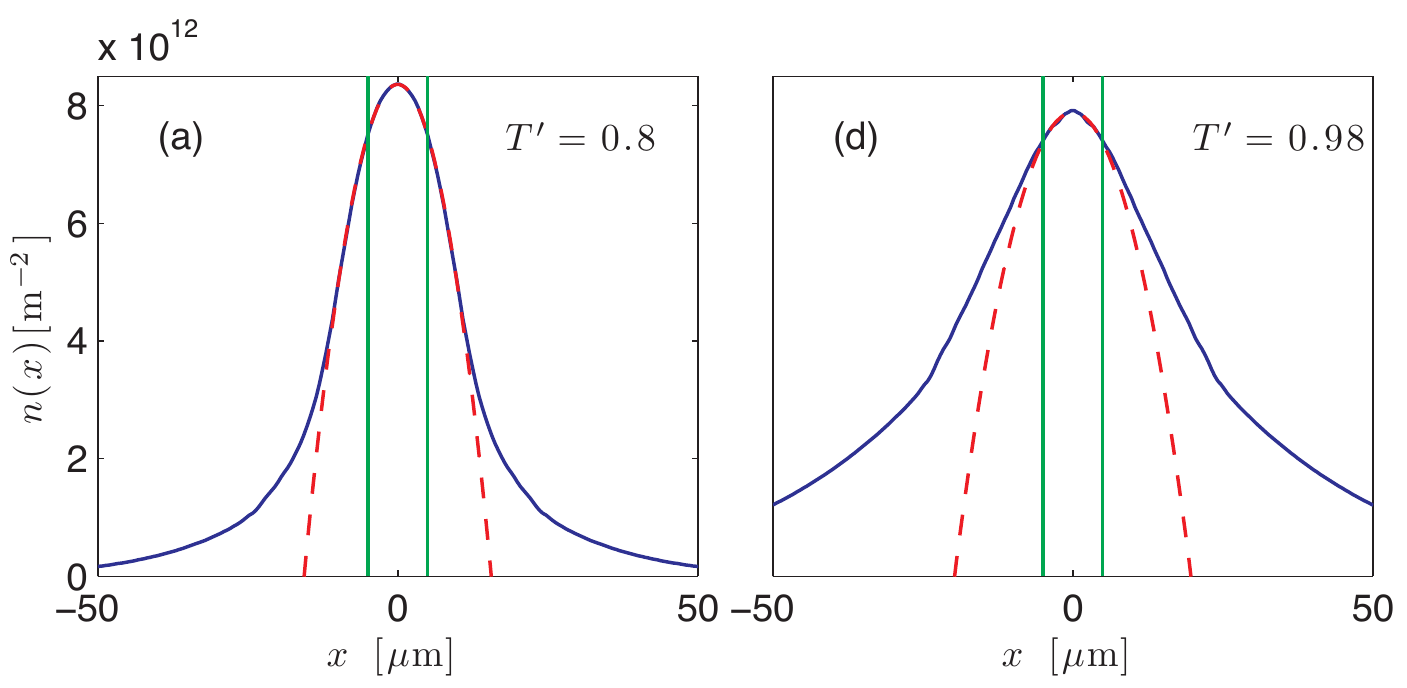} 
\caption{\label{Fig:cent_curve_fit_eg} Total areal density profiles at several temperatures: $n(x)$ (solid curve), fitting region (vertical lines indicate  $x=\pm5\mu$ m) and quadratic fits (dashed). Parameters as in Fig.~\ref{Fig:DegenFracs}.}
\end{figure}
Holzmann and Krauth  \cite{Holzmann2008} proposed using the central density curvature
\begin{equation}
\kappa=-\frac{\partial n(x)}{\partial(x^2)}\Big|_{x=0},
\end{equation}
as a signature of the superfluid transition, showing with QMC calculations that $\kappa$ was peaked near the superfluid transition. However they only calculated a small number of results across the transition region  and  it was not possible to clarify if this peak occurred at $T_c$ or $T_{BKT}$.

In Fig.~\ref{Fig:cent_curve_fit_eg} we show two examples of our c-field calculated average density profiles, and the
quadratic curves fitted to the central region which we use to extract the curvature. In general we observe that the quality of the fits become progressively poorer with increasing temperature. We find that the curvature evaluated this way  depends on the size of the fitting region, and we use $|x|\le5\mu\rm{m}$. 

\begin{figure}
\centering
\includegraphics[width=2.80in]{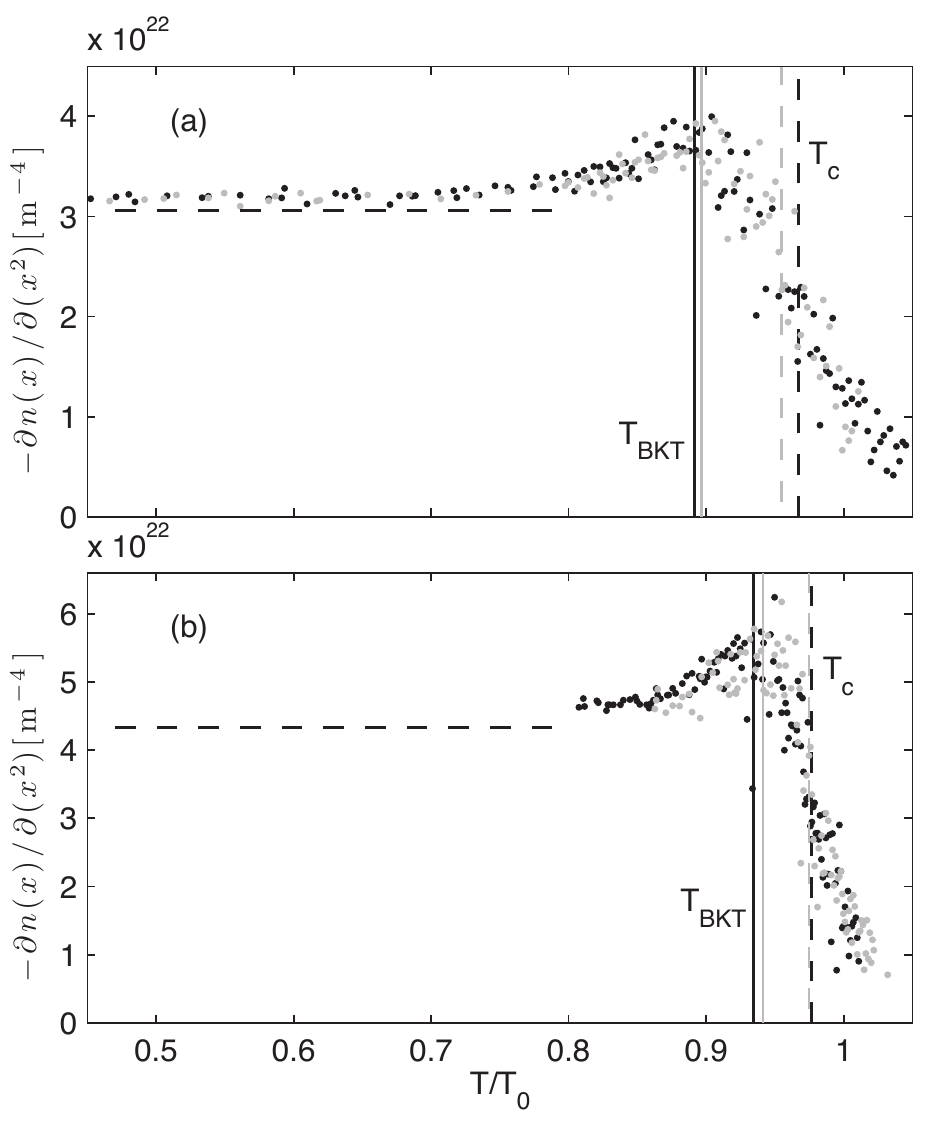}
\caption{\label{Fig:centralcurve5mum} Central density curvature. 
Parameters: (a) $(\omega_{x,y},\omega_z)=2\pi\times(9.4,1880)$ Hz with $\tilde{g}=0.107$, (black dots) system with $N=13.5\times10^3$ at $T_{BKT}=34.7$ nK, (gray dots) system with $N=21.2\times10^3$ at $T_{BKT}=43.1$ nK. (b) $(\omega_{x,y},\omega_z)=2\pi\times(9.4,940)$ Hz with $\tilde{g}=0.075$, (black dots) system with $N=30.2\times10^3$ at $T_{BKT}=46.8$ nK, (gray dots) system with $N=67.3\times10^3$ at $T_{BKT}=64.9$ nK. Vertical solid and dashed lines indicate $T_{BKT}$ and $T_c$, respectively. The Horizontal dashed line shows $\kappa_{\rm{TF}}$. }
\end{figure}

Figure \ref{Fig:centralcurve5mum} presents results for the central curvature as a function of temperature. 
We find, in qualitative agreement with the QMC results in Ref.~\cite{Holzmann2008}, that the curvature is peaked in the transition region.
For low temperatures ($T<T_{BKT}$) the curvature approaches a constant at a value slightly greater than that of the Thomas-Fermi profile ($\kappa_{\rm{TF}} =m^2\omega^2/2\hbar^2\tilde{g}$). This is consistent with the system density fluctuations being strongly suppressed at trap center.  Above the transition region ($T\sim T_{BKT}$) we observe the curvature decrease with increasing temperature.


The range of results in  Fig.~\ref{Fig:centralcurve5mum} span a considerable range of system parameters, and show that the curvature peak near $T_{BKT}$ is robust.  Our results also show that as the system size increases $T_{BKT}$ and $T_c$ gradually get closer (e.g.~compare the black and gray results in Fig.~\ref{Fig:centralcurve5mum}). However, our results do not allow us to conclude whether this is because of a reduction in the finite size effect (due to larger atom number) or because the system is crossing over to being three-dimensional (due to the increase in transition temperature relative to $\hbar\omega_z/k_B$).

\subsection{Condensate number fluctuations}
Recently the c-field technique has been extended to probe mode statistics in detail, and applied to exploring the critical exponents and emergence of coherence for a three-dimensional trapped Bose gas. Here we apply these techniques to the quasi-2D Bose gas.

\begin{figure}
\centering
\includegraphics[width=3.2in]{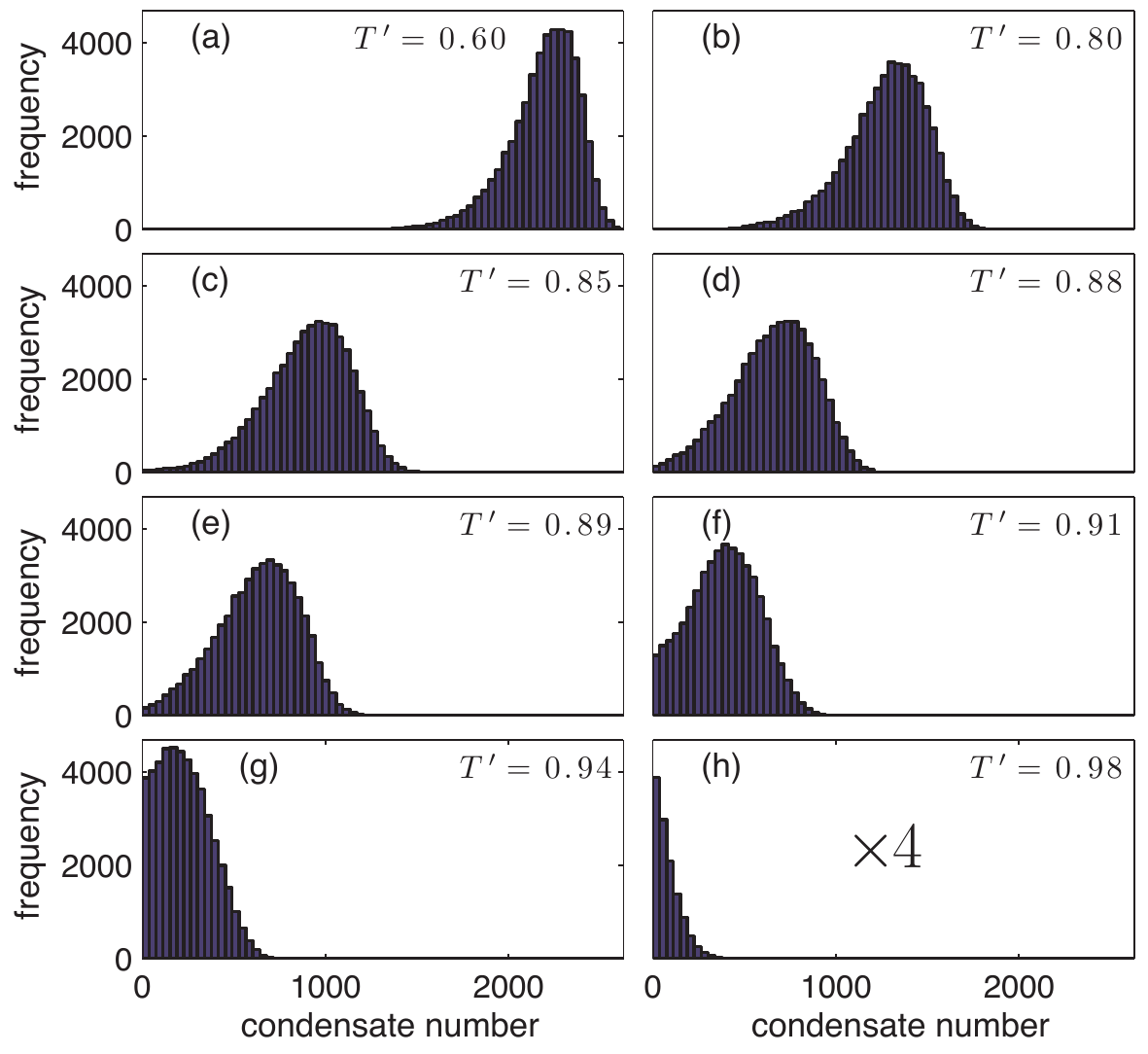}
\caption{\label{Fig:condFlucLowT}Condensate number distribution:  Histograms of the condensate number extracted from c-field evolutions at a range of temperatures. The "$\times 4$" indicates that the frequencies have been divided by 4 for subplot (h) for convenience. Parameters as in Fig.~\ref{Fig:DegenFracs}.}
\end{figure}

The procedure for extracting the condensate number statistics is discussed in 
detail in \cite{Bezett2009b}. Briefly, we obtain samples of the condensate amplitude ($\alpha_c(t)$) at time $t$, by taking the inner product of the classical field at this time with the (time-averaged) condensate mode\footnote{Taken as the eigenvector of the one-body density matrix corresponding to the largest eigenvalue, which above $T_c$ approximately corresponds to the harmonic oscillator ground mode.}. We can then construct the condensate number distribution, with number samples given by $N_c(t)=|\alpha_c(t)|^2$.  For each simulation we  evaluate  $5\times10^4$ samples of $N_c$ taken over  1000 radial trap periods of c-field evolution. Figure \ref{Fig:condFlucLowT}  shows the resulting histograms of the condensate number for eight different temperature regimes.

\begin{figure}
\centering
\includegraphics[width=3.2in]{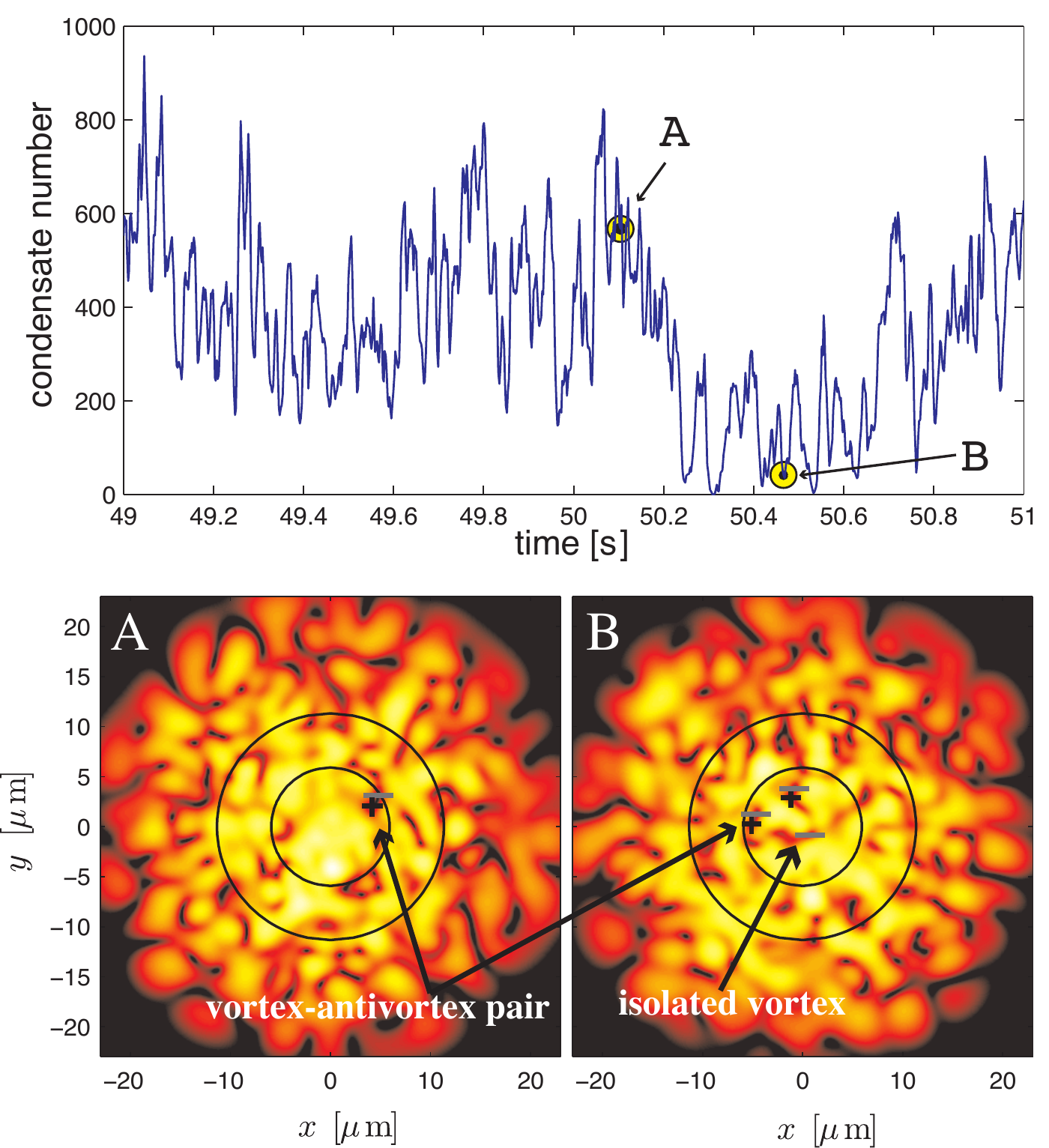}
\caption{\label{Fig:condFluc_t_TBKTzoom} (upper) Condensate number variation during c-field evolution for system at $T'=0.91$, i.e.~same data as for Fig.~\ref{Fig:condFlucLowT}(f). (lower) C-field density of the A,B microstates at the times indicated in the upper plot. Inner (outer) circle indicates the condensate (quasi-condensate) $1/e$ radius. Vortices (+) and antivortices (-) indicated inside the central condensate region only.
}
\end{figure}

The results at the  lowest [Fig.~\ref{Fig:condFlucLowT}(a)] and highest [Fig.~\ref{Fig:condFlucLowT}(h)] temperatures show number distributions consistent with coherent and incoherent number statistics, respectively.  
What is most intriguing is the qualitative change in behavior we observe in the temperature range from $T_{BKT}$ to $T_c$. For $T\sim T_{BKT}$  [Fig.~\ref{Fig:condFlucLowT}(e)] we find that the distribution is maximum at a macroscopic number ($N_{c,\max}\approx700$), with negligibly small probability of finding $N_c=0$. For temperatures between $T_{BKT}$ and $T_c$ the condensate number distribution has an interesting shape: the most likely condensate number is a macroscopic, non-zero value, but the probability of finding $N_c=0$ is appreciable (e.g.~see Figs.~\ref{Fig:condFlucLowT}(f) and \ref{Fig:condFlucLowT}(g)). 
In this regime condensate number fluctuations are large relative to the mean value. 
Figure \ref{Fig:condFluc_t_TBKTzoom} demonstrates a $N_c(t)$ trajectory from a c-field simulation for this regime. This trajectory reveals a typical event that occurs at this temperature: a sustained period ($50.2$ s $\lesssim t\lesssim 50.7$ s) over which $N_c$ is suppressed to a value significantly below the long time average. Examining the microstates we find that this suppression occurs when an isolated vortex enters the central region of the system (see microstate B in Fig.~\ref{Fig:condFluc_t_TBKTzoom}). These \textit{free vortices} are long lived and for $T_{BKT}<T<T_c$ they penetrate the central region quite frequently.   
Below $T_{BKT}$ such free vortices are strongly suppressed (near trap center) leading to a negligible probability of observing $N_c\approx0$ (of course this does not exclude paired vortices). 
The role of vortices was speculated to be the origin of fluctuations observed in the NIST experiment \cite{Clade2009} for temperature above $T_{BKT}$, although limited resolution prevented the direct observation of these vortices (also see\cite{Stock2005}).

\section{Conclusions}
 We have analyzed two important characteristics of the low-temperature trapped quasi-2D Bose gas. Our results for the central density curvature of the system show that this quantity peaks near the transition region, in qualitative agreement with previous quantum Monte Carlo work. The feature of our work is that our results resolve the transition region in detail and clearly reveal that the curvature peak occurs in the vicinity of $T_{BKT}$ (and not $T_c$), and is thus distinct from the momentum space bimodality arising at $T_c$. So far most experiments have examined the system density profile after expansion for intermediate times, where the imaged density is  a convolution of the \textit{in situ} position and momentum space distributions (e.g.~see \cite{Hadzibabic2008}). This suggests that the use of longer time-of-flights (or Bose-gas focusing \cite{Amerongen2008a}) to measure the momentum distribution, and \textit{in situ} imaging to directly observe the position density, will more clearly reveal the distinction between $T_c$ and $T_{BKT}$ in the quasi-two dimensional gas.
 
Our results for the condensate fluctuations help clarify the changes that occur when coherence and superfluidity emerge in the system. In particular, it reveals that the condensate above $T_{BKT}$ is frequently penetrated by free vortices causing strong number fluctuations.  These predictions broadly support the interpretation presented in Ref.~\cite{Clade2009}.  To date no direct experimental measurements of the condensate number statistics have been made. However, with the emergence of new techniques for accurately measuring critical properties \cite{Donner2007a}, and proposals for the use of related observables (e.g.~fluctuations in the zero momentum density \cite{Bezett2009b}) to infer this information, we expect such measurements will be feasible in the future.

{\bf Acknowledgments:}
PBB and RNB are supported by NZ-FRST contract NERF-UOOX0703.
RNB acknowledges support from the Otago Research Committee.


\begin{thebibliography}{24}
\expandafter\ifx\csname natexlab\endcsname\relax\def\natexlab#1{#1}\fi
\expandafter\ifx\csname bibnamefont\endcsname\relax
  \def\bibnamefont#1{#1}\fi
\expandafter\ifx\csname bibfnamefont\endcsname\relax
  \def\bibfnamefont#1{#1}\fi
\expandafter\ifx\csname citenamefont\endcsname\relax
  \def\citenamefont#1{#1}\fi
\expandafter\ifx\csname url\endcsname\relax
  \def\url#1{\texttt{#1}}\fi
\expandafter\ifx\csname urlprefix\endcsname\relax\def\urlprefix{URL }\fi
\providecommand{\bibinfo}[2]{#2}
\providecommand{\eprint}[2][]{\url{#2}}

\bibitem[{\citenamefont{Hadzibabic et~al.}(2006)\citenamefont{Hadzibabic,
  Kr{\"u}ger, Cheneau, Battelier, and Dalibard}}]{Hadzibabic2006}
\bibinfo{author}{\bibfnamefont{Z.}~\bibnamefont{Hadzibabic}},
  \bibinfo{author}{\bibfnamefont{P.}~\bibnamefont{Kr{\"u}ger}},
  \bibinfo{author}{\bibfnamefont{M.}~\bibnamefont{Cheneau}},
  \bibinfo{author}{\bibfnamefont{B.}~\bibnamefont{Battelier}},
  \bibnamefont{and} \bibinfo{author}{\bibfnamefont{J.}~\bibnamefont{Dalibard}},
  \bibinfo{journal}{Nature} \textbf{\bibinfo{volume}{441}},
  \bibinfo{pages}{1118} (\bibinfo{year}{2006}).

\bibitem[{\citenamefont{Kr\"{u}ger et~al.}(2007)\citenamefont{Kr\"{u}ger,
  Hadzibabic, and Dalibard}}]{Kruger2007}
\bibinfo{author}{\bibfnamefont{P.}~\bibnamefont{Kr\"{u}ger}},
  \bibinfo{author}{\bibfnamefont{Z.}~\bibnamefont{Hadzibabic}},
  \bibnamefont{and} \bibinfo{author}{\bibfnamefont{J.}~\bibnamefont{Dalibard}},
  \bibinfo{journal}{Phys. Rev. Lett.} \textbf{\bibinfo{volume}{99}},
  \bibinfo{eid}{040402} (\bibinfo{year}{2007}).

\bibitem[{\citenamefont{Clad\'{e} et~al.}(2009)\citenamefont{Clad\'{e}, Ryu,
  Ramanathan, Helmerson, and Phillips}}]{Clade2009}
\bibinfo{author}{\bibfnamefont{P.}~\bibnamefont{Clad\'{e}}},
  \bibinfo{author}{\bibfnamefont{C.}~\bibnamefont{Ryu}},
  \bibinfo{author}{\bibfnamefont{A.}~\bibnamefont{Ramanathan}},
  \bibinfo{author}{\bibfnamefont{K.}~\bibnamefont{Helmerson}},
  \bibnamefont{and} \bibinfo{author}{\bibfnamefont{W.~D.}
  \bibnamefont{Phillips}}, \bibinfo{journal}{Phys. Rev. Lett.}
  \textbf{\bibinfo{volume}{102}}, \bibinfo{pages}{170401}
  (\bibinfo{year}{2009}).

\bibitem[{\citenamefont{Simula and Blakie}(2006)}]{Simula2006}
\bibinfo{author}{\bibfnamefont{T.~P.} \bibnamefont{Simula}} \bibnamefont{and}
  \bibinfo{author}{\bibfnamefont{P.~B.} \bibnamefont{Blakie}},
  \bibinfo{journal}{Phys. Rev. Lett.} \textbf{\bibinfo{volume}{96}},
  \bibinfo{pages}{020404} (\bibinfo{year}{2006}).

\bibitem[{\citenamefont{Holzmann and Krauth}(2008)}]{Holzmann2008}
\bibinfo{author}{\bibfnamefont{M.}~\bibnamefont{Holzmann}} \bibnamefont{and}
  \bibinfo{author}{\bibfnamefont{W.}~\bibnamefont{Krauth}},
  \bibinfo{journal}{Phys. Rev. Lett.} \textbf{\bibinfo{volume}{100}},
  \bibinfo{pages}{190402} (\bibinfo{year}{2008}).

\bibitem[{\citenamefont{Bisset et~al.}(2009{\natexlab{a}})\citenamefont{Bisset,
  Davis, Simula, and Blakie}}]{Bisset2009}
\bibinfo{author}{\bibfnamefont{R.~N.} \bibnamefont{Bisset}},
  \bibinfo{author}{\bibfnamefont{M.~J.} \bibnamefont{Davis}},
  \bibinfo{author}{\bibfnamefont{T.~P.} \bibnamefont{Simula}},
  \bibnamefont{and} \bibinfo{author}{\bibfnamefont{P.~B.}
  \bibnamefont{Blakie}}, \bibinfo{journal}{Phys. Rev. A}
  \textbf{\bibinfo{volume}{79}}, \bibinfo{pages}{033626}
  (\bibinfo{year}{2009}{\natexlab{a}}).

\bibitem[{\citenamefont{Giorgetti et~al.}(2007)\citenamefont{Giorgetti,
  Carusotto, and Castin}}]{Giorgetti2007a}
\bibinfo{author}{\bibfnamefont{L.}~\bibnamefont{Giorgetti}},
  \bibinfo{author}{\bibfnamefont{I.}~\bibnamefont{Carusotto}},
  \bibnamefont{and} \bibinfo{author}{\bibfnamefont{Y.}~\bibnamefont{Castin}},
  \bibinfo{journal}{Phys. Rev. A} \textbf{\bibinfo{volume}{76}},
  \bibinfo{pages}{013613} (\bibinfo{year}{2007}).

\bibitem[{\citenamefont{Bezett and Blakie}(2009)}]{Bezett2009b}
\bibinfo{author}{\bibfnamefont{A.}~\bibnamefont{Bezett}} \bibnamefont{and}
  \bibinfo{author}{\bibfnamefont{P.~B.} \bibnamefont{Blakie}},
  \bibinfo{journal}{Phys. Rev. A} \textbf{\bibinfo{volume}{79}},
  \bibinfo{eid}{033611} (\bibinfo{year}{2009}).

\bibitem[{\citenamefont{Petrov et~al.}(2000)\citenamefont{Petrov, Holzmann, and
  Shlyapnikov}}]{Petrov2000}
\bibinfo{author}{\bibfnamefont{D.~S.} \bibnamefont{Petrov}},
  \bibinfo{author}{\bibfnamefont{M.}~\bibnamefont{Holzmann}}, \bibnamefont{and}
  \bibinfo{author}{\bibfnamefont{G.~V.} \bibnamefont{Shlyapnikov}},
  \bibinfo{journal}{Phys. Rev. Lett.} \textbf{\bibinfo{volume}{84}},
  \bibinfo{pages}{2551} (\bibinfo{year}{2000}).

\bibitem[{\citenamefont{Stock et~al.}(2005)\citenamefont{Stock, Hadzibabic,
  Battelier, Cheneau, and Dalibard}}]{Stock2005}
\bibinfo{author}{\bibfnamefont{S.}~\bibnamefont{Stock}},
  \bibinfo{author}{\bibfnamefont{Z.}~\bibnamefont{Hadzibabic}},
  \bibinfo{author}{\bibfnamefont{B.}~\bibnamefont{Battelier}},
  \bibinfo{author}{\bibfnamefont{M.}~\bibnamefont{Cheneau}}, \bibnamefont{and}
  \bibinfo{author}{\bibfnamefont{J.}~\bibnamefont{Dalibard}},
  \bibinfo{journal}{Phys. Rev. Lett.} \textbf{\bibinfo{volume}{95}},
  \bibinfo{pages}{190403} (\bibinfo{year}{2005}).

\bibitem[{\citenamefont{Blakie and Davis}(2007)}]{Blakie2007CR}
\bibinfo{author}{\bibfnamefont{P.~B.} \bibnamefont{Blakie}} \bibnamefont{and}
  \bibinfo{author}{\bibfnamefont{M.~J.} \bibnamefont{Davis}},
  \bibinfo{journal}{J. Phys. B} \textbf{\bibinfo{volume}{40}},
  \bibinfo{pages}{2043} (\bibinfo{year}{2007}).

\bibitem[{\citenamefont{Davis et~al.}(2001)\citenamefont{Davis, Ballagh, and
  Burnett}}]{Davis2001}
\bibinfo{author}{\bibfnamefont{M.~J.} \bibnamefont{Davis}},
  \bibinfo{author}{\bibfnamefont{R.~J.} \bibnamefont{Ballagh}},
  \bibnamefont{and} \bibinfo{author}{\bibfnamefont{K.}~\bibnamefont{Burnett}},
  \bibinfo{journal}{J. Phys. B: Atomic} \textbf{\bibinfo{volume}{34}},
  \bibinfo{pages}{4487} (\bibinfo{year}{2001}).

\bibitem[{\citenamefont{Blakie and Davis}(2005)}]{Blakie2005}
\bibinfo{author}{\bibfnamefont{P.~B.} \bibnamefont{Blakie}} \bibnamefont{and}
  \bibinfo{author}{\bibfnamefont{M.~J.} \bibnamefont{Davis}},
  \bibinfo{journal}{Phys. Rev. A} \textbf{\bibinfo{volume}{72}},
  \bibinfo{pages}{063608} (\bibinfo{year}{2005}).

\bibitem[{\citenamefont{Blakie et~al.}(2008)\citenamefont{Blakie, Bradley,
  Davis, Ballagh, and Gardiner}}]{Blakie2008}
\bibinfo{author}{\bibfnamefont{P.~B.} \bibnamefont{Blakie}},
  \bibinfo{author}{\bibfnamefont{A.~S.} \bibnamefont{Bradley}},
  \bibinfo{author}{\bibfnamefont{M.~J.} \bibnamefont{Davis}},
  \bibinfo{author}{\bibfnamefont{R.~J.} \bibnamefont{Ballagh}},
  \bibnamefont{and} \bibinfo{author}{\bibfnamefont{C.~W.}
  \bibnamefont{Gardiner}}, \bibinfo{journal}{Adv. Phys.}
  \textbf{\bibinfo{volume}{57}}, \bibinfo{pages}{363} (\bibinfo{year}{2008}).

\bibitem[{\citenamefont{Simula et~al.}(2008)\citenamefont{Simula, Davis, and
  Blakie}}]{Simula2008}
\bibinfo{author}{\bibfnamefont{T.~P.} \bibnamefont{Simula}},
  \bibinfo{author}{\bibfnamefont{M.~J.} \bibnamefont{Davis}}, \bibnamefont{and}
  \bibinfo{author}{\bibfnamefont{P.~B.} \bibnamefont{Blakie}},
  \bibinfo{journal}{Phys. Rev. A} \textbf{\bibinfo{volume}{77}},
  \bibinfo{pages}{023618} (\bibinfo{year}{2008}).

\bibitem[{\citenamefont{Sato et~al.}(2009)\citenamefont{Sato, Suzuki, and
  Kawashima}}]{Sato2009}
\bibinfo{author}{\bibfnamefont{T.}~\bibnamefont{Sato}},
  \bibinfo{author}{\bibfnamefont{T.}~\bibnamefont{Suzuki}}, \bibnamefont{and}
  \bibinfo{author}{\bibfnamefont{N.}~\bibnamefont{Kawashima}},
  \bibinfo{journal}{J. Phys.: Conf. Ser.} \textbf{\bibinfo{volume}{150}},
  \bibinfo{pages}{032094} (\bibinfo{year}{2009}).

\bibitem[{\citenamefont{Prokof'ev et~al.}(2001)\citenamefont{Prokof'ev,
  Ruebenacker, and Svistunov}}]{Prokofev2001}
\bibinfo{author}{\bibfnamefont{N.}~\bibnamefont{Prokof'ev}},
  \bibinfo{author}{\bibfnamefont{O.}~\bibnamefont{Ruebenacker}},
  \bibnamefont{and}
  \bibinfo{author}{\bibfnamefont{B.}~\bibnamefont{Svistunov}},
  \bibinfo{journal}{Phys. Rev. Lett.} \textbf{\bibinfo{volume}{87}},
  \bibinfo{pages}{270402} (\bibinfo{year}{2001}).

\bibitem[{\citenamefont{Bagnato and Kleppner}(1991)}]{Bagnato1991}
\bibinfo{author}{\bibfnamefont{V.}~\bibnamefont{Bagnato}} \bibnamefont{and}
  \bibinfo{author}{\bibfnamefont{D.}~\bibnamefont{Kleppner}},
  \bibinfo{journal}{Phys. Rev. A} \textbf{\bibinfo{volume}{44}},
  \bibinfo{pages}{7439} (\bibinfo{year}{1991}).

\bibitem[{\citenamefont{Prokof\char39{}ev and Svistunov}(2002)}]{Prokofev2002}
\bibinfo{author}{\bibfnamefont{N.}~\bibnamefont{Prokof\char39{}ev}}
  \bibnamefont{and}
  \bibinfo{author}{\bibfnamefont{B.}~\bibnamefont{Svistunov}},
  \bibinfo{journal}{Phys. Rev. A} \textbf{\bibinfo{volume}{66}},
  \bibinfo{pages}{043608} (\bibinfo{year}{2002}).

\bibitem[{\citenamefont{Holzmann et~al.}(2008)\citenamefont{Holzmann,
  Chevallier, and Krauth}}]{Holzmann2008A}
\bibinfo{author}{\bibfnamefont{M.}~\bibnamefont{Holzmann}},
  \bibinfo{author}{\bibfnamefont{M.}~\bibnamefont{Chevallier}},
  \bibnamefont{and} \bibinfo{author}{\bibfnamefont{W.}~\bibnamefont{Krauth}},
  \bibinfo{journal}{Europhys. Lett} \textbf{\bibinfo{volume}{82}},
  \bibinfo{pages}{30001} (\bibinfo{year}{2008}).

\bibitem[{\citenamefont{Bisset et~al.}(2009{\natexlab{b}})\citenamefont{Bisset,
  Baillie, and Blakie}}]{Bisset2009A}
\bibinfo{author}{\bibfnamefont{R.~N.} \bibnamefont{Bisset}},
  \bibinfo{author}{\bibfnamefont{D.}~\bibnamefont{Baillie}}, \bibnamefont{and}
  \bibinfo{author}{\bibfnamefont{P.~B.} \bibnamefont{Blakie}},
  \bibinfo{journal}{Phys. Rev. A} \textbf{\bibinfo{volume}{79}},
  \bibinfo{pages}{013602} (\bibinfo{year}{2009}{\natexlab{b}}).

\bibitem[{\citenamefont{Hadzibabic et~al.}(2008)\citenamefont{Hadzibabic,
  Krüger, Cheneau, Rath, and Dalibard}}]{Hadzibabic2008}
\bibinfo{author}{\bibfnamefont{Z.}~\bibnamefont{Hadzibabic}},
  \bibinfo{author}{\bibfnamefont{P.}~\bibnamefont{Krüger}},
  \bibinfo{author}{\bibfnamefont{M.}~\bibnamefont{Cheneau}},
  \bibinfo{author}{\bibfnamefont{S.~P.} \bibnamefont{Rath}}, \bibnamefont{and}
  \bibinfo{author}{\bibfnamefont{J.}~\bibnamefont{Dalibard}},
  \bibinfo{journal}{New J. Phys.} \textbf{\bibinfo{volume}{10}},
  \bibinfo{pages}{045006} (\bibinfo{year}{2008}).

\bibitem[{\citenamefont{van Amerongen et~al.}(2008)\citenamefont{van Amerongen,
  van Es, Wicke, Kheruntsyan, and van Druten}}]{Amerongen2008a}
\bibinfo{author}{\bibfnamefont{A.~H.} \bibnamefont{van Amerongen}},
  \bibinfo{author}{\bibfnamefont{J.~J.~P.} \bibnamefont{van Es}},
  \bibinfo{author}{\bibfnamefont{P.}~\bibnamefont{Wicke}},
  \bibinfo{author}{\bibfnamefont{K.~V.} \bibnamefont{Kheruntsyan}},
  \bibnamefont{and} \bibinfo{author}{\bibfnamefont{N.~J.} \bibnamefont{van
  Druten}}, \bibinfo{journal}{Phys. Rev. Lett.} \textbf{\bibinfo{volume}{100}},
  \bibinfo{eid}{090402} (\bibinfo{year}{2008}).

\bibitem[{\citenamefont{Donner et~al.}(2007)\citenamefont{Donner, Ritter,
  Bourdel, Ottl, Kohl, and Esslinger}}]{Donner2007a}
\bibinfo{author}{\bibfnamefont{T.}~\bibnamefont{Donner}},
  \bibinfo{author}{\bibfnamefont{S.}~\bibnamefont{Ritter}},
  \bibinfo{author}{\bibfnamefont{T.}~\bibnamefont{Bourdel}},
  \bibinfo{author}{\bibfnamefont{A.}~\bibnamefont{Ottl}},
  \bibinfo{author}{\bibfnamefont{M.}~\bibnamefont{Kohl}}, \bibnamefont{and}
  \bibinfo{author}{\bibfnamefont{T.}~\bibnamefont{Esslinger}},
  \bibinfo{journal}{Science} \textbf{\bibinfo{volume}{315}},
  \bibinfo{pages}{1556} (\bibinfo{year}{2007}).

\end{thebibliography}
\end{document}